\documentstyle[12pt]{article}
\def\R{{\rm I\hspace{-.15em}R}}

\def\1{\mbox{I\hspace{-.15em}1}}
\def\C{\hspace{3pt}{\rm l\hspace{-.47em}C}}
\def\ba{\begin{array}}
\def\ea{\end{array}}
\catcode `\@=11 \@addtoreset{equation}{section}
\def\theequation{\arabic{section}.\arabic{equation}}

\def\1{\mathbf {Id} }

\def\b{\begin{equation}}
\def\e{\end{equation}}

\begin{document}

\title{\textbf{``Massless'' Spin-$\frac{3}{2}$ Fields in the de Sitter Space}}
\author{ A. Azizi$^{1}$ \thanks{e-mail: aazizi@iausdj.ac.ir- alissr2010@gmail.com}, M. Amiri$^{2}$}

\maketitle

\centerline{\it $^1$ Department of Physics,
Sanandaj Branch, Islamic Azad University, Sanandaj,
IRAN }\centerline{\it $^2$Department of Physics, Razi University,
Kermanshah, IRAN}\vspace{15pt}

\begin{abstract}

In this paper, the ``massless" spin-$\frac{3}{2}$ fields in the de
Sitter space are considered. This work is in the continuation of
a previous paper devoted to the quantization of the de Sitter
``massive" spin-$\frac{3}{2}$ fields. Due to the appearance of gauge
invariance and indefinite metric, the covariant quantization of the ``massless"
spin-$\frac{3}{2}$ fields requires an indecomposable representation
of the de Sitter group. The gauge fixing corresponding to the simplest Gupta-Bleuler
structure is used, and gauge invariant field is discussed. The field equation is obtained by using the Casimir operator of the de Sitter group. The solutions are written in terms of the coordinate-independent de Sitter plane waves. Finally, the generalized two-point function is calculated.
\end{abstract}


\vspace{0.5cm} {\it Proposed PACS numbers}: 04.62.+v, 98.80.Cq,
12.10.Dm \vspace{0.5cm} 

\newpage
\section{Introduction}

In the previous work \cite{az}, the ``massive'' spin-$\frac{3}{2}$
fields in the de Sitter space have been studied. The term ``massive" refers to those the de Sitter fields which
unambiguously contract to the Minkowskian massive fields at zero curvature limit. These fields are
associated with the principal series of unitary irreducible
representations of the de Sitter group $SO_0(1,4)$, with the Casimir operator
eigenvalue $<Q_\nu^{(1)}>=\nu^2-\frac{3}{2},\;\;\nu
> \frac{3}{2}$ (or equivalently ``mass''
$m_p^2=H^2(\nu^2-\frac{3}{2}))$. The interpretation of $m_p$ as a
mass becomes made possible by examining its null curvature limit. The quotation marks on ``mass''
are used because of the absence of the intrinsic notion of mass in
the de Sitter relativity \cite{gano}. Indeed, the principal series of unitary irreducible
representations admits a massive representation of the Poincar\'e
group in the limit $H=0$ \cite{pot,eric,mini}.

In this paper, the ``massless" spin-$\frac{3}{2}$ fields in the de Sitter space are considered. ``Massless" is used by a
reference to the conformal invariance and propagation on the light-cone. They are associated with the discrete series of unitary irreducible representations of the de Sitter group. There are two unitary irreducible
representations of the de Sitter group in the discrete series with a Minkowskian meaning in the null curvature limit, which are denoted by
$\Pi^{\pm}_{\frac{3}{2},\frac{3}{2}}$ with the same the Casimir operator eigenvalue  $<Q^{(1)}>=-\frac{5}{2}$.

The field equation of the ``massless" spin-$\frac{3}{2}$ fields is gauge invariant in
the de Sitter space as well as the massless fields in the Minkowski space for $s\geq1$.
It is well known that the quantization of gauge invariant theories
usually requires the \emph{\`a la} Gupta-Bleuler quantization
\cite{binegar,ga}. It has been proved that the use of an indefinite
metric is an unavoidable feature if one insists on the preserving of
causality (locality) and covariance in gauge quantum field theories
\cite{st}. This means that  one can not limit the state space of
the massless fields (with $s\geq 1$) to Hilbert space, the
appearance of states with negative or null norms is necessary for
the quantization of gauge fields. Therefore, an indecomposable structure appears inevitable, where the physical states belong to a subspace $V$ of the solutions, but where the field operator must be defined on a larger gauge dependent
space $V_\lambda$ (which contains negative norm states). The
physical subspace $V$ is invariant but not invariantly complemented
in $V_\lambda$. The same feature is repeated in $V$, where one can
find the invariant (but again not invariantly complemented) subspace
of gauge solutions $V_g$. These gauge solutions have zero norms and
are orthogonal to all the elements of $V$  \cite{gahamu}.
Consequently, one must eliminate them from the subspace $V$  by
considering the physical state space as the coset $V/V_g$.

For simplicity, the following units are used: $$c=\hbar=1,\:\:\:\
[x^\alpha/H]=1,\:\:\:\ [M]=H,$$ where $c$, $\hbar$ and $H$ are light
velocity, Planck constant and Hubble parameter respectively. The paper is organized as follows: In section 2, some of the useful notations of de Sitter space and unitary irreducible representations of the de Sitter group will be recalled.
Section $3$ is devoted to derivation the de Sitter-spin-$\frac{3}{2}$
``massless" field equation. In this section, we will show that the
field equation is gauge invariant. We will adopt a very convenient
value for the gauge fixing parameter $\lambda$. In this paper, we
get the second order wave equation by using the Casimir operator,
then it is converted to the first order equation. There are two
different equations for this case; one of those is considered in the context
and the other is studied in Appendix. In section $4$, the solutions are calculated in terms of the coordinate-independent de Sitter plane waves. It will be shown that, for $\lambda=\frac{1}{2}$, the field solution has a simple form. In section
$5$, we will define the two-point function $S(x,y)$ that satisfies
the following conditions: a) indefinite sesquilinear form, b)
locality, c) covariance and d) normal analyticity. Normal
analyticity allows us to define the two-point function $S(x,y)$ as the
boundary value of the analytic function $\texttt{S}(z_1,z_2)$ from
the tube domains. The normal analyticity is related to the Hadamard
condition that selects a unique vacuum state. $\texttt{S}(z_1,z_2)$
is defined in terms of the spinor-vector de Sitter plane-waves in their tube
domains. Section $6$ contains a brief conclusion and the outlook.

\section{The de Sitter space notations}

The de Sitter space is visualized as the hyperboloid with equation:
$$ X_H=\{x \in \R^5;x^2=\eta_{\alpha\beta} x^\alpha     x^\beta
=-H^{-2}=-\frac{3}{\Lambda}\},\;
$$ where $\eta_{\alpha\beta}=\mbox{diag}(1,-1,-1,-1,-1);
\alpha,\beta=0,1,...,4,$ and $\Lambda$ is a positive cosmological
constant. The metric is
$$ds^{2}=\eta_{\alpha\beta}dx^{\alpha}dx^{\beta}\vert_{x^2=-H^{-2}}=g_{\mu\nu}^{dS}dX^{\mu}dX^{\nu};\;\;\mu,\nu=0,1,2,3,$$
where $X^\mu$ are the four local space-time coordinates on de Sitter
hyperboloid. A spinor-tensor field $\Psi_{\alpha_1...\alpha_l}(x)$
on $X_H$ can be viewed as an homogeneous function on $ \R^{5}$
variables $x^\alpha$ with an arbitrary degree of homogeneity
$\sigma$. Also it must satisfy the following conditions \cite{dir}:
$$x\cdot\partial \Psi=\sigma \Psi,\:\:\:\:\:\:\:\:\:\:(homogeneity),$$
$$x\cdot \Psi(x)=0,\:\:\:\:\:\:\:\:\:\:(transversality).$$
On de Sitter space the tangential (or transverse) derivative defines as
\begin{equation}
\partial^\top_\alpha=\theta_{\alpha\beta}\partial^\beta=\partial_\alpha
+H^2x_\alpha x\cdot\partial,\ \ \ \ x\cdot\partial^\top=0,
\end{equation}
where $\theta_{\alpha\beta}=\eta_{\alpha\beta}+H^2x_{\alpha}x_{
\beta}$ is transverse projection tensor ($\theta_{\alpha\beta}\;
x^{\alpha}=\theta_{\alpha\beta} \; x^{\beta}=0$).

The unitary irreducible
representations of $10$-parameter group SO$_0(1,4)$ (connected component of
the identity) of the de Sitter space, which is one of the two possible
deformations of the Poincar\'e group (the other one being
SO$_0(2,3)$), are characterized by the eigenvalues of the two
Casimir operators $Q^{(1)}$ and $Q^{(2)}$. These operators  that
commute with the group generators  are constant in each unitary irreducible
representation. They read
      \b Q^{(1)}=-\frac{1}{2}L_{\alpha\beta}L^{\alpha\beta}\:\:\:\:,\:\:\:\:Q^{(2)}=-W_\alpha W^\alpha\;\;\;,\;\;\;W_\alpha =\frac{1}{8}
      \epsilon_{\alpha\beta\gamma\delta\eta} L^{\beta\gamma}L^{\delta\eta},\e
where $\epsilon_{\alpha\beta\gamma\delta\eta}$ is the usual
antisymmetrical tensor in $\R^5$ and
$L_{\alpha\beta}=M_{\alpha\beta}+S_{\alpha\beta}$ is an infinitesimal
generator. The orbital part $M_{\alpha\beta}$ is
          \b  M_{\alpha \beta}=-i(x_\alpha \partial_\beta-x_\beta
      \partial_\alpha)=-i(x_\alpha\partial^\top_\beta-x_\beta
        \partial^\top_\alpha).\e
In order to precise the action of the spinorial  part
$S_{\alpha\beta}$ on a field tensor or spinor-tensor one must treat
separately the integer and half-integer cases. Tensor fields of rank
$l$, $\Psi_{\gamma_1...\gamma_l}(x)$, show integer spin fields, and
the spinorial action is \cite{ga}
       \b S_{\alpha \beta}^{(l)}\Psi_{\gamma_1...\gamma_l}=-i\sum^l_{i=1}
          \left(\eta_{\alpha\gamma_i}
        \Psi_{\gamma_1...(\gamma_i\rightarrow\beta)... \gamma_l}-\eta_{\beta\gamma_i}
          \Psi_{\gamma_1...(\gamma_i\rightarrow \alpha)... \gamma_l}\right),\e
where $(\gamma_i\rightarrow\beta)$ means $\gamma_i$ index replaced
with $\beta$. Half-integer spin fields with spin $s=l+\frac{1}{2}$
are represented by four component spinor-tensor $\Psi_{
\gamma_{1}...\gamma_{l} }^{i}$ with spinor index $i=1,2,3,4$. The
spinorial part now reads
$$S_{\alpha\beta}^{(s)}=S_{\alpha\beta}^{(l)}+S_{\alpha\beta}^{(\frac{1}{2})},\qquad
\mbox{with}\qquad
S_{\alpha\beta}^{(\frac{1}{2})}=-\frac{i}{4}\left[\gamma_{\alpha},\gamma_{\beta}\right],$$
and with the Dirac gamma matrices $\gamma_{\alpha}$
\cite{TAK2,TAK3,BOG}
$$\gamma^{\alpha}\gamma^{\beta}+\gamma^{\beta}\gamma^{\alpha}
=2\eta^{\alpha\beta},\qquad
\gamma^{\alpha\dagger}=\gamma^{0}\gamma^{\alpha}\gamma^{0}\;,$$the
useful representations, which are compatible with group as follows
$$ \gamma^0=\left( \begin{array}{clcr} I & \;\;0 \\ 0 &-I \\ \end{array} \right)
      ,\gamma^4=\left( \begin{array}{clcr} 0 & I \\ -I &0 \\ \end{array} \right) , $$ \b
   \gamma^1=\left( \begin{array}{clcr} 0 & i\sigma^1 \\ i\sigma^1 &0 \\
    \end{array} \right)
   ,\gamma^2=\left( \begin{array}{clcr} 0 & -i\sigma^2 \\ -i\sigma^2 &0 \\
      \end{array} \right)
   , \gamma^3=\left( \begin{array}{clcr} 0 & i\sigma^3 \\ i\sigma^3 &0 \\
      \end{array} \right),\e
where $\sigma_i$ are Pauli matrices and $I$ is a $2\times2$ unit
matrix. The Casimir operators are simple to manipulate in ambient
space notation. Since $Q^{(1)}$ is a second order derivative
operator, it is convenint used for obtaining the field equation. In
particular, it is easy to show that for a $l$-rank tensor field
$\Psi_{\gamma_{1}...\gamma_{l}}(x)$ one has
 \b Q_l^{(1)}\Psi=Q_0^{(1)}\Psi-2\Sigma_1 \partial x .\Psi+2\Sigma_1 x \partial.
           \Psi+2\Sigma_2 \eta \Psi'-l(l+1)\Psi,\e
where
   \b Q_l^{(1)}=-\frac{1}{2}L_{\alpha \beta}^{(l)}L^{\alpha \beta(l)}
         =-\frac{1}{2}M_{\alpha \beta}M^{\alpha \beta}-\frac{1}{2} S_{\alpha
           \beta}^{(l)}S^{\alpha
            \beta(l)} -M_{\alpha \beta}S^{\alpha \beta(l)},\e

\b M_{\alpha \beta}S^{\alpha \beta(l)}\Psi(x)= 2\Sigma_1
\partial x
       .\Psi-2\Sigma_1 x \partial.\Psi-2l\Psi,\e
       \b\frac{1}{2}S_{\alpha \beta}^{(l)}S^{\alpha \beta(l)}\Psi=l(l+3)\Psi-
          2\Sigma_2 \eta \Psi',\e

\b Q_0^{(1)}=-\frac{1}{2}M_{\alpha \beta}M^{\alpha \beta},\e $\Psi'$
is the trace of the $l$-rank tensor $\Psi(x)$ and $\Sigma_p$ is the
non-normalized symmetrization operator:
         \b \Psi'_{\alpha_1...\alpha_{l-2}}=\eta^{\alpha_{l-1}\alpha_l}
          \Psi_{\alpha_1...\alpha_{l-2} \alpha_{l-1}\alpha_l},\e
         \b (\Sigma_p AB)_{\alpha_1...\alpha_l}=\sum_{i_1<i_2<...<i_p}
          A_{\alpha_{i_1}\alpha_{i_2}...\alpha_{i_p}}
          B_{\alpha_1...\not\alpha_{i_1}...\not\alpha_{i_2} ...\not\alpha_{i_p}...\alpha_l}.\e

For half-integer spin fields with spin $s=l+\frac{1}{2}$, the
$S_{\alpha\beta}^{(\frac{1}{2})}$ acts only upon the index i, and we
have \cite{BOG,LESM}
$${\cal S}^{(\frac{1}{2})}_{\alpha\beta}{\cal S}^{\alpha\beta(l)}\Psi(x)=l \Psi(x)-\Sigma_1\gamma(\gamma\cdot
\Psi(x)).$$ In this case, the Casimir operator is
          $$ Q^{(1)}_s=-\frac{1}{2}\left(M_{\alpha \beta}+S_{\alpha \beta}^{(l)}+
                S_{\alpha\beta}^{(\frac{1}{2})}\right)
      \left(M^{\alpha \beta}+S^{\alpha \beta(l)}+S^{\alpha \beta(\frac{1}{2})}\right)$$
   \b =Q^{(1)}_l-\frac{5}{2}+\frac{i}{2}\gamma_{\alpha}\gamma_{\beta}M^{\alpha
\beta}-
             S_{\alpha \beta}^{(\frac{1}{2})}S^{\alpha \beta(l)}.\e
Then we obtain
       \b Q^{(1)}_s\Psi(x)=\left(Q^{(1)}_l-l-\frac{5}{2}+\frac{i}{2}\gamma_{\alpha}\gamma_{\beta}M^{\alpha
       \beta}\right)\Psi(x)+ \Sigma_1 \gamma (\gamma.\Psi(x)) ,\e
or
      $$ Q^{(1)}_s\Psi(x)=\left(-\frac{1}{2}M_{\alpha \beta}M^{\alpha
      \beta}+\frac{i}{2} \gamma_{\alpha}\gamma_{\beta}
       M^{\alpha \beta}-l(l+2)-\frac{5}{2}\right)\Psi(x)$$ \b -2\Sigma_1
         \partial x .\Psi(x)+2\Sigma_1 x \partial.\Psi(x)+2\Sigma_2 \eta
       \Psi'(x)+\Sigma_1 \gamma (\gamma.\Psi(x)).\e
As you will see in the next section, the spin-$\frac{3}{2}$ field
equation can be written in terms of the Casimir operator
$Q_{}^{(1)}$ .

\section{Field equation and Gauge transformation}

\subsection{Field equation}

As previously mentioned, the operator $Q_{\frac{3}{2}}^{(1)}$
commutes with the group generators and consequently it is constant
on each unitary irreducible
representation. In fact, we can classify the spinor-vector unitary irreducible
representations by using the eigenvalues of $Q^{(1)}$. The field
equation can be written as
\b\left(Q_{\frac{3}{2}}^{(1)}-<Q_{\frac{3}{2}}^{(1)}>\right)\Psi(x)=0.\label{3.1}
\e In Takahashi \cite{TAKA} and Dixmier \cite{DIX},  a general
classification scheme for all the unitary irreducible
representations of the de Sitter group, is expressed and may
be labeled by a pair of parameters $(p,q)$ with $2p \in N$ and $q
\in C$ as follows
\begin{equation}
<Q^{(1)}>=[-p(p+1)-(q+1)(q-2)], \quad <Q^{(2)}>=[-p(p+1)q(q-1)].
\label{casimirs}
\end{equation}
 According to the possible values of
$p$ and $q$, two types of unitary irreducible
representations of the spin-$\frac{3}{2}$ field are
distinguished for de Sitter group $SO(1,4)$ namely, the principal and the discrete series. More particularity about mathematical details of
the group contraction and the relationship between the de Sitter and the
Poincar\'e groups are given in the \cite{LE,BAC}. The unitary irreducible representations of the spin-$\frac{3}{2}$ field relevant to the present work are as
follows:
\begin{itemize}
\item[i)] The unitary irreducible representations
$U^{\frac{3}{2},\nu}$ in the principal series where $p = s =
\frac{3}{2}$ and $q = \frac{1}{2} + i\nu$ matching to the Casimir
spectral values:$$
<Q_\frac{3}{2}^{(1)}>=\nu^2-\frac{3}{2},\;\;\;\;\nu \in \R \:\:\:
\nu>\frac{3}{2}.
$$
Note that $U^{\frac{3}{2},\nu}$ and $U^{\frac{3}{2},-\nu}$ are
equivalent.
\item[ ii)]  The unitary irreducible representations $\Pi^{\pm}_{\frac{3}{2},q}$ of the discrete
series, where $p = s = \frac{3}{2}$, correspond to  \b
<Q_\frac{3}{2}^{(1)}>= -\frac{5}{2} ,\; \;\;\;
q=\frac{3}{2},\:\:\:\:\:\Pi^{\pm}_{\frac{3}{2},\frac{3}{2}}, \label{3.3}\e \b
<Q_\frac{3}{2}^{(1)}>= -\frac{3}{2} ,\; \;\;\;
q=\frac{1}{2},\:\:\:\:\:\Pi^{\pm}_{\frac{3}{2},\frac{1}{2}}.\e
\end{itemize}

The physical content of the principal series and the discrete series
representation from the point of view of a Minkowskian observer at
the limit $H=0$ have been expressed in \cite{az}.

The ``massless'' spin-$\frac{3}{2}$ field in the de Sitter space corresponds to
the discrete series $\Pi^\pm_{\frac{3}{2},\frac{3}{2}}$ and the
field equation is
\begin{equation}
\left(Q_\frac{3}{2}^{(1)}+\frac{5}{2}\right)\Psi(x)=0,
\label{dirac1}
\end{equation} where \b Q^{(1)}_{\frac{3}{2}}\Psi(x)=\left(-\frac{1}{2}M_{\alpha \beta}M^{\alpha
      \beta}+\frac{i}{2} \gamma_{\alpha}\gamma_{\beta}
       M^{\alpha \beta}-3-\frac{5}{2}\right)\Psi(x)$$ $$ -2\partial x.\Psi(x)+2x\partial.\Psi(x)+\gamma
       \gamma.\Psi(x).\e
If we compare this equation with the massive spin-$\frac{3}{2}$
field  case (equation (3.7) in \cite{az}) they are equivalent for
the values  $\nu=\pm i$ but these values do not admissible for the
unitary irreducible
representations of the de Sitter group in the principal series. It is natural to use the
solution of the equation (\ref{3.1}), which are already given for the
massive case \cite{az}. The corresponding spin-$\frac{3}{2}$ field
solution can be written under the form \cite{az}:
\begin{equation}
\Psi_{\alpha}(x)=\frac{1}{\langle
Q_{\frac{3}{2}}^{(1)}\rangle+\frac{5}{2}} {\cal E}_\alpha
(x,\xi)\psi (x)\,,
\end{equation}
where ${\cal E}_\alpha (x,\xi)$, and $\psi (x)$  also contain
constant terms involving the parameters $p$ and $q$, where do not
diverge for the specific values $p=q=\frac{3}{2}$ corresponding to
the relation (\ref{3.3}). One can simply see that the field solution
for these values is singular (equation (4.12) in \cite{az}). Clearly,
the singularity appears only for the spin-$\frac{3}{2}$ massless field
for representations ${\Pi}^{\pm}_{\frac{3}{2},\frac{3}{2}}$.  This
singularity appears due to the gauge invariance, which will be
discussed in the next subsection. Therefore the subspace
$\partial.\Psi=0$ alone can not be used for the construction of a
quantum massless spin-$\frac{3}{2}$ field. One must solve the
equation in a larger space, which includes the
$\partial\cdot\Psi\neq 0$ types of solutions. As expected, one finds
three main types of solutions: the general solutions which aren't
divergenceless, the divergencelessness type and the gauge type
solution.
\subsection{Gauge transformation}

The ``massless'' spin-$\frac{3}{2}$ field, however with the
subsidiary conditions $\partial^\top. \Psi=0=\gamma .\Psi(x)$, is
singular. This type of singularity is actually due to the
divergencelessness condition needed to associate this field with a
specific unitary irreducible
representation of the de Sitter group. To solve this problem, the subsidiary
conditions must be dropped. Then the field equation (\ref{dirac1}) replaced
with the following equation
 \b (Q^{(1)}_{\frac{3}{2}}+\frac{5}{2}) \Psi_\alpha(x)-D_{\frac{3}{2}\alpha} \partial^\top.\Psi(x)=0,\e
where $
D_{\frac{3}{2}\alpha}=-H^{-2}\partial^\top_\alpha-\gamma^\top_\alpha\not
x$,
$\not x=\gamma _\alpha x^\alpha$ and
$\gamma^\top_\alpha=\theta_{\alpha\beta}\gamma^\beta$. One can show
that this equation is invariant under gauge transformation \b
\Psi_\alpha(x)\rightarrow
\Psi'_\alpha(x)=\Psi_\alpha(x)+D_{\frac{3}{2}\alpha}\zeta ,\e where
$\zeta$ is an arbitrary spinor field. For providing gauge
invariant the following identities are used:
$$Q^{(1)}_{\frac{3}{2}}D_{\frac{3}{2}}=D_{\frac{3}{2}}Q^{(1)}_{\frac{1}{2}}\:\:\:,\:\:\:\partial^\top.D_{\frac{3}{2} }\zeta=Q^{(1)}_{\frac{1}{2}}\zeta+
\frac{5}{2}\zeta.$$ Let us introduce a gauge fixing parameter
$\lambda$. The wave equation now reads as \b
\left(Q_\frac{3}{2}^{(1)}+\frac{5}{2}\right)\Psi(x)-\lambda
D_{\frac{3}{2}\alpha}
\partial^\top.\Psi(x)=0,\label{3.10}\e the role of $\lambda$ is just to fix the gauge spinor field $\zeta$.

It is convenient and usual to continue this work with the first
order field equation for the spinor case. We can write the equation
(\ref{dirac1}) such as
 \b \tilde{D}^\beta_\alpha D^\delta_\beta \Psi_\delta=0,\:\:\:\:\:\: or\:\:\:\:\:\:
D^\delta_\beta\tilde{D}^\beta_\alpha\Psi_\delta=0, \e where
$\tilde{D}^\beta_\alpha$ and $D^\delta_\beta $ are defined by \b
\left\{ \ba{rcl} \tilde{D}^\delta_\beta =\not
x\not\partial^\top\eta^\delta_\beta-3\eta^\delta_\beta-x_\beta\not
x\gamma^\delta,
\\
D^\beta_\alpha=\not
x\not\partial^\top\eta^\beta_\alpha-\eta^\beta_\alpha-x_\alpha\not
x\gamma^\beta. \ea\right.\e There are two possibilities for the
first order field equation as follows:\b \tilde
D^\delta_\alpha\Psi_\delta=0, \label{3.13}\e\b
D^\delta_\alpha\Psi_\delta=0,\label{3.14} \e where for physical
states, means $\lambda=0$ in equation(\ref{3.10}), both first-order
equation(\ref{3.13}) and (\ref{3.14}) are equivalent, since their
multiplication gives the equation(\ref{dirac1}). But for unphysical
states(dependent gauge states,i.e. $\lambda\neq0$ and pure gauge
states,i.e. $\lambda=1$), the solutions of the two equations are
different, although the physical parts or the central parts are
equal in these cases. The equation (\ref{3.13}) is considered in
Appendix A.1. Here we consider equation (\ref{3.14}). This equation
can be written as a simple form \b \not
x\not\partial^\top\Psi_\alpha(x)-x_\alpha\not
x\not\Psi-\Psi_\alpha(x)=0 .\e  For prettiness it can be written as
\b \left(Trpr\not x\not\partial^\top-1\right)\Psi_\alpha(x)=0
,\label{3.16}\e where $(Trpr)$ means is the transverse projection.
We should be recalled that the massless spin-$\frac{3}{2}$ is
singular when $\nu=\pm i$. This singularity refereed to gauge
invariant as we have this difficult for all fields with spin s
$\geq1$. Now we rewrite the equation (\ref{3.16}) as\b
\left(Trpr\not
x\not\partial^\top-1\right)\Psi_\alpha(x)+D_{\frac{3}{2}\alpha}\not
x\not\Psi(x)=0 ,\e which is invariant under gauge
transformation: \b \Psi_\alpha (x) \longrightarrow \Psi_\alpha'
(x)=\Psi_\alpha (x)+
\partial^\top_\alpha\zeta ,\label{3.18}\e  $\zeta$ is an arbitrary
spinor field. Let us introduce a gauge fixing parameter $\lambda$
for this case. Now the wave equation reads as \b \left(Trpr\not
x\not\partial^\top-1\right)\Psi_\alpha(x)+\lambda
D_{\frac{3}{2}\alpha}\not x\not\Psi(x)=0 .\label{3.19}\e The above
equation under gauge transformation (\ref{3.18}) becomes \b
\left(Trpr\not x\not\partial^\top-1\right)\Psi_\alpha(x)+\lambda
D_{\frac{3}{2}\alpha}\not
x\not\Psi(x)=(1-\lambda)D_{\frac{3}{2}\alpha}\not
x\not\partial^\top\zeta. \e With the choice of value $\lambda\neq
1$, the gauge field $\partial^\top_\alpha\zeta$ is put into the
field equation only if $\zeta$ obeys \b(Trpr\not
x\not\partial^\top-1)\partial^\top_\alpha\zeta=
-D_{\frac{3}{2}\alpha}\not x\not\partial^\top\zeta=0,\e that means
the gauge field $\zeta$ is fixed. The adjoint $\Psi_\alpha(x)$ is
defined as follows \cite{TAK2,BOG} $$\bar\Psi_\alpha(x)\equiv
\Psi^{\dag}_\alpha(x){\gamma^0}{\gamma^4},$$ which satisfy the
equation \b
\bar\Psi_\alpha\gamma^4(\overleftarrow{\not\partial}^\top\not
x)\gamma^4-\bar\Psi_\beta\gamma^4(\gamma^\beta x_\alpha\not
x)\gamma^4+\bar\Psi_\alpha+c\bar\Psi_\beta\gamma^4\gamma^\beta\not
x\overleftarrow{D}_{\frac{3}{2}\alpha}\gamma^4=0. \label{3.24}\e

As stated in \cite{binegar}, ``the appearance of (the Gupta-Bleuler)
triplet seems to be universal in gauge theories, and crucial for
quantization". The ambient space formalism will allow to exhibit
this triplet for the present field in exactly the same manner as it
occurs for the electromagnetic field.

Let us now define the Gupta-Bleuler triplet $V_g \subset V \subset
V_{\lambda}$ carrying the indecomposable structure for the unitary irreducible
representations of the de Sitter
group appearing in our problem:

\begin{itemize}

\item[-] The space  $V_{\lambda}$ is the space  of all square integrable
 solutions of the field equation (\ref{3.19}). It is $\lambda$ dependent so that one can
actually adopt an optimal value of $\lambda$ which has very simple
form. In the next section, we will show  that this value is
$\lambda=\frac{1}{2}$, (more generally for a spin s field, we have
$\lambda=(2/(2s+1))$ \cite{ga}).

\item[-] It contains a closed subspace $V$ of solutions satisfying
the conditions $\partial^\top\cdot\Psi=0=\gamma\cdot\Psi(x)$. This invariant subspace $V$ is not
invariantly complemented in $V_\lambda$. In view of equation
(\ref{3.19}), it is obviously $\lambda$ independent.

\item[-]The subspace $V_g$ of $V$ consists of the gauge solutions
of the form $\partial^\top_\alpha\zeta$, and
$\not\partial^\top\zeta=0=Q_0\zeta$. These are orthogonal to every
element in $ V$ including themselves. They form an invariant
subspace of $V$ but admit no invariant complement in $V$.
\end{itemize}
The de Sitter group acts on the physical  (or transverse) space $V/V_{g}$
through the massless, helicity $\pm \frac{3}{2}$ unitary
representation $ \Pi^+_{\frac{3}{2},\frac{3}{2}} \bigoplus
\Pi^-_{\frac{3}{2},\frac{3}{2}}$, which are called the central parts
in this paper.

\section{The de Sitter-Spin-$\frac{3}{2}$ Plane Waves}

According to the de Sitter plane waves, which were presented by Bros et al.
\cite{BROS1}, the de Sitter-Dirac plane wave for spinor field was
calculated\cite{BOG}. In this section, the spinor-vector solution
can be written by using de Sitter-Dirac plane wave as terms in the
following form
 \b \Psi_\alpha (x)= Z^\top_\alpha \psi_1+D_{\frac{3}{2}\alpha} \psi_2+\gamma^\top_\alpha\psi_3,\label{4.1} \e
where $Z$ is an arbitrary five-component constant vector field: $$
Z^\top_\alpha =\theta_{\alpha\beta} Z^\beta=Z_\alpha+H^2x_\alpha
x\cdot Z,\;x\cdot Z^\top=0.$$ By putting $\Psi_\alpha$ in equation
(\ref{3.19}), we find that the spinor fields $\psi_1,\psi_2$, and
$\psi_3$ must obey the following equations: \b (\not
x\not\partial^\top-1)\psi_1=0,\label{39}\e  \b \not
x\not\partial^\top\psi_3+\not x(\not
x\not\partial^\top-4)\psi_2=\not xx.Z\psi_1,\label{4.3}\e \b
2(1-2\lambda)\psi_3+\not x\left[(1-\lambda)\not
x\not\partial^\top-2(1-2\lambda)\right] \psi_2=\lambda\not
Z^\top\psi_1.\label{4.4}\e For $\lambda=\frac{1}{2}$ equation
(\ref{4.4}) becomes \b \not\partial^\top\psi_2=-\not Z^\top\psi_1
.\label{4.5}\e By multiplying equation (\ref{39}) in $\not
\partial^\top$, we have
$$Q_0\psi_1=2\psi_1.$$
 Since the spinor fields $\psi_1$,$\psi_2$, and
$\psi_3$ are homogeneous function of variate $x$ with the same
degree of homogeneity, therefor by using equation (\ref{4.5}), we
obtain \b Q_0\psi_2=\left(4x.Z+2Z.\not\partial^\top+\not Z^\top\not
x\right)\psi_1,\e  $\psi_2$ can be written in the following form:\b
\psi_2=Q_0^{-1}\left(4x.Z+2Z.\partial^\top+\not Z^\top\not
x\right)\psi_1+\psi_g,\label{4.7}\e where $\psi_g$ is $\psi_g=\phi_g\textbf{U}$ and $Q_0\phi_g=0.$ $\textbf{U}$ is an
arbitrary constant spinor field, and $\phi_g$ is a massless
minimally coupled scalar field. By using the following identities:
\b Q_0x.Z\psi_1=-2x.Z\psi_1-2Z.\partial^\top\psi_1,\e \b
Q_0Z.\partial^\top\psi_1=4x.Z\psi_1+4Z.\partial^\top\psi_1,\e \b
Q_0\not Z^\top\not x\psi_1=2x.Z\psi_1+2Z.\partial^\top\psi_1,\e and
\b Q_{0}\left(4x.Z+2Z.\partial^\top+\not Z^\top\not
x\right)\psi_1=Q_0\not Z^\top\not x\psi_1=2\not Z^\top\not
x\psi_1,\e we obtain\b \psi_2=\frac{1}{2}\not Z^\top\not
x\psi_1+\psi_g.\label{4.12}\e By replacing equation (\ref{4.12}) in
equation (\ref{4.3}), we obtain (see Appendix (A.2))\b \psi_3=\not
x\psi_g.\label{4.13}\e If equations (\ref{4.12}) and (\ref{4.13})
are substituted in equation (\ref{4.1}), it can be seen that in this
choice of gauge $\lambda=\frac{1}{2}$, the spinor field $\psi_g$ or
equivalently $\psi_3$ is removed, and the solution can be written in
the following simplest form \b \Psi_\alpha(x)={\cal
D}_\alpha\psi_1,\label{419}\e where \b {\cal
D}_\alpha=\frac{1}{2}\left[3Z^\top_\alpha+x.Z\partial^\top_\alpha-\gamma^\top_\alpha\not
Z^\top\right] ,\e and $\psi_1$ is the solution of the de Sitter-Dirac
field equation. In the previous paper, the spinor field $\psi_1$ was
explicitly calculated, and the solutions are given by \cite{TAK2,BOG}
\b (\psi_1)_1={\cal V}(x,\xi)(Hx.\xi )^{-3}, \e \b (\psi_1)_2={\cal
U}(\xi)(Hx.\xi)^{-1},\e where ${\cal V}(x,\xi)=\not x \not \xi{\cal
V}(\xi)$ and
$$\xi \in {\cal C}^+=\{ \xi \;\;;\eta_{\alpha \beta}\xi^\alpha
\xi^\beta=(\xi^0)^2-\vec \xi.\vec \xi-(\xi^4)^2=0,\; \xi^0>0 \}.$$
The two spinors ${\cal V}(\xi)$ and ${\cal U}(\xi)$ are \b {\cal
U}^a(\xi)=\frac{\xi^0-\vec \xi. \vec \gamma
 \gamma^0+1}{\sqrt{2(\xi^0+1)}}{\cal U}^a(\stackrel{o}{\xi}_+),\;\;\; {\cal V}^a(\xi)=\frac{1}{\sqrt{2(\xi^0+1)}}{\cal U}^a(\stackrel{o}{\xi}_-), \;\;a=1,2, \e where
  \b {\cal U}_1(\stackrel{o}{\xi}_+) =\frac{1}{\sqrt{2}}\left( \begin{array}{clcr} \alpha\\ \alpha \\ \end{array} \right) ,\; {\cal U}_2             (\stackrel{o}{\xi}_+)=\frac{1}{\sqrt{2}}\left( \begin{array}{clcr} \beta\\ \beta \\ \end{array} \right),\e
\b {\cal U}_1(\stackrel{o}{\xi}_-) =\frac{1}{\sqrt{2}}\left( \begin{array}{clcr} \alpha\\-\alpha \\ \end{array} \right) ,\; {\cal U}_2             (\stackrel{o}{\xi}_-)=\frac{1}{\sqrt{2}}\left( \begin{array}{clcr} \beta\\-\beta \\ \end{array} \right),\e
with $\alpha=\left( \begin{array}{clcr}1 \\ 0 \\ \end{array} \right)$ , $\beta=\left( \begin{array}{clcr} 0 \\ 1 \\ \end{array} \right)$ and $ \xi =\stackrel{o}{\xi}_\pm \equiv (1,\vec 0,\pm 1)$.  Finally the two possible solutions for
$\Psi_\alpha(x)$ are
$$\Psi^a_{1\alpha}(x)=\frac{1}{2}\left[3Z^\top_\alpha+x.Z\partial^\top_\alpha-\gamma^\top_\alpha\not
Z^\top\right]\frac{\not x \not \xi}{x.\xi}{\cal
V}(\xi) (Hx.\xi )^{-2}$$\b\equiv{\cal V_\alpha}(x,\xi,Z)(Hx.\xi )^{-2} ,\e
 and
$$\Psi^a_{2\alpha}(x)=\frac{1}{2}\left[3Z^\top_\alpha+x.Z\partial^\top_\alpha-\gamma^\top_\alpha\not
Z^\top\right]{\cal U}(\xi)(Hx.\xi)^{-1}$$\b \equiv{\cal
U_\alpha}(x,\xi,Z)(Hx.\xi)^{-1} .\e By taking the derivation of the plane
waves $(x.\xi)^\sigma,$ the explicit forms of ${\cal U}_\alpha$ and
${\cal V}_\alpha$ are obtained in term of $\xi$ as follows  \b
{\cal
V}_\alpha(x,\xi,Z)=\frac{1}{2}\left[3Z^\top_\alpha-3\frac{x.Z}{x.\xi}\xi^\top_\alpha-x.Z\gamma^\top_\alpha-\gamma^\top_\alpha\not
Z^\top\right]\frac{\not x \not \xi}{x.\xi}{\cal V}(\xi) ,\e and\b
{\cal
U}_\alpha(x,\xi,Z)=\frac{1}{2}\left[3Z^\top_\alpha-\frac{x.Z}{x.\xi}\xi^\top_\alpha-\gamma^\top_\alpha\not
Z^\top\right]{\cal U}(\xi) .\e the spinor field $\psi_1$ satisfy the
field equation \b (Q_0-2)\psi_1=0.\label{62}\e It is corresponds to
the massless conformally invariant field equation \cite{BOG,BR}.

The arbitrariness introduced with the constant vector $Z$ will be
removed  by comparison the solution with the Minkowskian limit.
Unfortunately, our notations for the ``massless'' conformally
coupled scalar field are not adapted to the computation of the limit
$H=0$ . It is due to the fact that contrary to the ``massive'' case
the values $\sigma=-1,-2$ are constant \cite{brepmo}. In order to
obtain the behavior of the field solutions in the limit $H=0$ (at
least for the scalar part), one can use the global conformal
coordinate system
$$ x_H(X)=(x^0=H^{-1}\sinh HX^0, \vec x=H^{-1} \frac{\vec
X}{\parallel \vec X\parallel} \cosh HX^0 \sin H\parallel \vec
X\parallel,$$ \b x^4= H^{-1}  \cosh HX^0 \cos H\parallel \vec
X\parallel),\e where $X^0=\rho, X^1=\alpha, X^2=\theta, X^3=\phi$
\cite{GARI2}. The square-integrable solutions of the field equation
are given by \cite{chta}
\begin{equation}\label{plane} \phi(x)=\phi(\rho, \vec v)=\cos{\rho}\,
\frac{e^{\pm i(L+1)\rho}}{\sqrt{L+1}}\,
\mathrm{Y}_{Llm}({\vec v}),
\end{equation}
where $\mathrm{Y}_{Llm}(\vec{v})$ are the  hyperspherical harmonics
on $S^3=\{ v^i \in R^4 | v_1^2+v_2^2+v_3^2+v_4^2=r^2\}$. The unitary irreducible
representations
$\Pi^{\pm}_{\frac{3}{2},\frac{3}{2}}$ of the de Sitter group correspond to the
unitary irreducible representations of the Poincar\'e group, in the limit H=0. Since, in this limit the
wave solutions are equivalent to the solutions of the Minkowskian space,
the numbers of the polarization states can be fixed in the same way
as in the Minkowskian counterpart. It can be shown that in the $H
=0\, , L \to \infty$ limit and with
\begin{equation}\label{limit}
\rho=Ht,\;\; \alpha=Hr;\;\; HL=k_0=|\vec
k|,\qquad\mbox{with}\quad\theta\,,\varphi\quad\mbox{unchanged}\,,\end{equation}
the functions (4.26) become, when suitably rescaled, the usual
massless spherical waves (with $k^2=(k^0)^2-(\vec k)^2=0$)
\cite{GARI}. With these coordinates, the de Sitter spinor-vector field
solutions read
\begin{equation}
\label{plane1} \Psi_{\mu}(\rho,\vec v)= V^{-1}(1\pm i\gamma^4)\frac{\partial
x^\alpha}{\partial
 X^\mu }\frac{1}{2}\left[3Z^\top_\alpha+x.Z\partial^\top_\alpha-\gamma^\top_\alpha\not
Z^\top\right] \cos \rho \frac{e^{\pm i(L+1)\rho}}{\sqrt{L+1}}
\mathrm{Y}_{Llm}(\\vec v)\,,
\end{equation}
where V is a spinor transformation matrix from ambient space notation to intrinsic coordinate which is defined in \cite{S.Moradi}.

The plane waves $(x\cdot\xi)^\sigma$ are singular at $x\cdot\xi=0$ and
they are not globally defined due to the ambiguity concerning to the
phase factor. In contrast with the Minkowskian exponentials plane wave,
these waves are singular on three-dimensional light-like manifolds
and can at first be defined only on suitable halves of $X_H$. We
will need an appropriate $i\epsilon$-prescription (indicated below)
to obtain global waves, for detail see \cite{BR}. For a complete determination, one may consider the
solution in the complex de Sitter  space-time $X_H^{(c)}$. The complex de Sitter
space-time is defined as \cite{BR}
$$ X_H^{(c)}=\left\{ z=x+iy\in  \C^5;\;\;\eta_{\alpha \beta}z^\alpha z^\beta=(z^0)^2-\vec z.\vec z-(z^4)^2=-H^{-2}\right\}$$
\b =\left\{ (x,y)\in  \R^5\times  \R^5;\;\; x^2-y^2=-H^{-2},\;
x.y=0\right\},\e where $T^\pm= \R^5+iV^\pm$ and the $V^+$(resp.
$V^-)$ stems from the causal structure on $X_H$: \b V^\pm=\left\{
x\in \R^5;\;\; x^0\stackrel{>}{<} \sqrt {\parallel \vec
x\parallel^2+(x^4)^2} \right\}.\e We introduce their respective
intersections with $X_H^{(c)}$, \b {\cal T}^\pm=T^\pm\cap
X_H^{(c)}.\e which will be called forward and backward tubes of the
complex de Sitter  space $X_H^{(c)}$. Then we define the ``tuboid'' above
$X_H^{(c)}\times X_H^{(c)}$ by  \b {\cal T}_{12}=\{ (z,z');\;\; z\in
{\cal T}^+,z' \in {\cal T}^- \}. \e More details are given in
\cite{BR}. When $z$ varies in ${\cal T}^+$ (or ${\cal T}^-$) and
$\xi$ lies in the positive cone ${\cal C}^+$:
$$\xi \in {\cal C}^+=\{ \xi \in {\cal C}; \; \xi^0>0 \}.$$
The sign of the imaginary part of $(z.\xi)$ has  fixed ,so the plane
wave solutions are globally defined. The phase is chosen such as \b
\mbox{boundary value of} \; (z.\xi)^\sigma \mid_{x.\xi>0}>0.\e
Finally we have \b \Psi_{1\alpha}(z)={\cal
U}_{\alpha}^{(\lambda)}(z,\xi)(Hz\cdot \xi)^{-1},\e
\begin{equation} \Psi_{2\alpha}(z)={\cal V}_{\alpha}^{(\lambda)}(z,\xi)
(Hz\cdot \xi)^{-3},\e where $z \in X_H^{(c)} $ and $\xi \in {\cal
C}^+$ these solutions are globally defiend in the de Sitter  hyperboloid and
they are independant of the choise of the intrisic coordinate.

\section{Two-point function}

The two-point function of ``massless''
spin-$\frac{3}{2}$ field is defined as \b S^{i \bar
j}_{\alpha\alpha'}(x,x')=<\Omega\mid\Psi^i_\alpha(x){\overline
\Psi}^{\overline j}_{\alpha'}(x')\mid\Omega>,\e where $x,x'\epsilon
X_H$. This function is a solution of the wave equation
$(\ref{3.19})$ with respect to $x$ and the wave equation
(\ref{3.24}) with respect to $x'$. In Appendix A.3, the solution of
equation (\ref{3.24}) is obtained and here we consider the equation
(\ref{3.19}). The solution can be found in terms of the two-point
function of spinor field, which was calculated in the previous paper
\cite{BOG}.

By using the recurrence formula (\ref{4.1}), we define\b S_{\alpha
\alpha'}(x,x')=\theta_{\alpha}. \theta'_{\alpha' }
S_1(x,x')-D_{\frac{3}{2}\alpha}
S_2(x,x'){\gamma^4}\overleftarrow{D}'_{\frac{3}{2}\alpha'}{\gamma^4}
-\gamma^\top_\alpha
S_3(x,x'){\gamma^4}{\gamma'}^\top_{\alpha'}{\gamma^4}.\e By imposing
the two-point function $S_{\alpha \alpha'}$ to obey equation
(\ref{3.19}) and by using the identities of equations
(\ref{39})-(\ref{4.4}), $S_1, S_2,$ and $S_3$ must be satisfied by
the following equations: \b (\not
x\not\partial^\top-1)S_1(x,x')=0,\label{5.3}\e \b \not
x\not\partial^\top{\cal
S}_3(x,x'){\gamma^4}\gamma'^\top_{\alpha'}{\gamma^4}+\not x(\not
x\not\partial^\top-4)S_2(x,x'){\gamma^4}\overleftarrow{D}'_{\frac{3}{2}\alpha'}{\gamma^4}=\not
xx.\theta'S_1(x,x'),\e $$
2(1-2\lambda)S_3(x,x'){\gamma^4}\gamma'^\top_{\alpha'}{\gamma^4}+\not
x\left[(1-\lambda)\not x\not\partial^\top-2(1-2\lambda)\right]
S_2(x,x'){\gamma^4}\overleftarrow{D}'_{\frac{3}{2}\alpha'}{\gamma^4}$$\b=\lambda
\gamma^\top.\theta'S_1(x,x').\label{5.5}\e For
$\lambda=\frac{1}{2},$ equation (\ref{5.5}) becomes \b
\not\partial^\top
S_2(x,x'){\gamma^4}\overleftarrow{D}'_{\frac{3}{2}\alpha'}{\gamma^4}=\gamma^\top.\theta'S_1(x,x'),\e
and we know that  $S_1$ is the solution of equation (\ref{5.3}) and
it is given by \cite{BOG} \b S_1(x,x')=\frac{1}{4\pi}\left[
3P^{(7)}_{-1}(x.x')\not x -P^{(7)}_{-3}(x.x')\not x'\right]
\gamma^{4}. \e  $S_2$ and $S_3$ are given by $$
S_2(x,x'){\gamma^4}\overleftarrow{D}'_{\frac{3}{2}\alpha'}{\gamma^4}=-\frac{1}{2}x.\theta'S_1+S_g,
$$\b S_3(x,x'){\gamma^4}{\gamma'}^\top_{\alpha'}{\gamma^4}=-\not xS_g.\e Finally the two-point function is obtained similar to
equation(\ref{419}) as
\begin{equation} S_{\alpha \alpha'}(x,x')=
D_{\alpha \alpha'}(x, \partial^\top ;x', {\partial'}^\top)
S_1(x,x'),\end{equation} where \b D_{\alpha \alpha'}=
\frac{1}{2}\left[3\theta_\alpha \cdot
\theta'_{\alpha'}+x\cdot\theta'_{\alpha}\partial^\top_{\alpha}
+\gamma^\top_{\alpha}\gamma\cdot\theta'\right],\e this function
satisfies the following conditions.
\begin{enumerate}
\item[a)] {\bf Indefinite sesquilinear form}: For any spinor-vector test function $f_\alpha \in {\cal D}(X_H)$, we have an
indefinite sesquilinear form that is defined by
\begin{equation} \int _{X_H \times X_H}
\bar f^{\alpha}_i(x)S^{i \bar j}_{\alpha \alpha'}(x,x') f^{\alpha'}_
{\bar j}(x')d\sigma(x)d\sigma(x'),\end{equation} where $ \bar f$ is
the adjoint of $f$, and $d\sigma (x)$ denotes the de Sitter -invariant
measure on $X_H$ \cite{BR}. ${\cal D}(X_H)$ is the space of
$C^\infty$ spinor-vector functions with compact support in $X_H$ and
with values in $\C^5$.

\item[b)] {\bf Covariance}
\b \Lambda^\alpha_\beta
\Lambda^{\alpha'}_{\beta'}g^{-1}S_{\alpha\alpha'}\bigl(\Lambda(g)x,\Lambda(g)x'\bigr)i(g)=S_{\beta\beta'}(x,x'),
\label{covarianza} \e where  $\Lambda \in SO_0(1,4)$ , $g\in
Sp(2,2)$ and $ g\gamma^\alpha g^{-1}=\Lambda_\beta^\alpha
\gamma^\beta$. $i(g)$ is the group involution defined by
\begin{equation}
i(g)=-\gamma^{4}g\gamma^{4}. \label{invol}
\end{equation}
\item[c)] {\bf Locality}: for every
space-like separated pair $(x,x')$, {\it i.e.} $x\cdot x'>-H^{-2}$,
\begin{equation} S_{\alpha \alpha'}^{i\overline{j}}(x,x')=-S_{ \alpha'
\alpha}^{\overline{j}i}(x',x),\end{equation}
where $S^{ \bar
j i}_{\alpha'\alpha}(x',x)=<\Omega\mid {\overline
\Psi}^{\overline j}_{\alpha'}(x')\Psi^i_\alpha(x)\mid\Omega>$.

\item[d)] {\bf Transversality}:

\begin{equation} x\cdot S(x,x')=0=x'\cdot S(x,x').\end{equation}

\item[e)] {\bf Normal analyticity}:
$S_{\alpha \alpha' }(x,x')$ is the boundary value (in the distributional sense ) of
an analytic function $\texttt{S}_{\alpha \alpha'}(z,z')$.

$\texttt{S}_{\alpha \alpha'}(z,z')$ is maximally analytic, i.e., can be analytically continued to the ``cut domain'' \cite{BOG,BR}:
$$\Delta=\{ (z,z') \in X_H^{(c)} \times X_H^{(c)} \;\; :\;\; (z-z')^2\leq 0 \}. $$
The two-point function $S_{\alpha \alpha' }(x,x')$ is the boundary
value of $\texttt{S}_{\alpha \alpha'}(z,z')$ from ${\cal T}_{12}$
and the ``permuted two-point function'' $S_{\alpha' \alpha }(x',x)$
is the boundary value of $\texttt{S}_{\alpha \alpha'}(z,z')$ from
the domain:
$$ {\cal T}_{21}=\{ (z,z');\;\; z'\in {\cal T}^+,z \in {\cal T}^- \}. $$

\end{enumerate}

\section{ Conclusions}

We have studied the ``massless'' spin-$\frac{3}{2}$ fields in the de Sitter
space-time in the ambient space formalism. This formalism is independent from the chosen coordinate
system. Gauge and conformal invariant are the properties of ``massless'' fields. Gauge invariant and
the Gupta-Bleuler triplet, are discussed. We have shown that the field equation of ``massless" spin-$\frac{3}{2}$ fields is gauge invariant. The two-point
function is calculated.  We saw that the spinor field $\psi_1$ in equation (\ref{62}) is conformal invariant. Conformal invariant of the spinor-vector field will be considered in the forthcoming paper \cite{gesh}.

\vspace{3mm} \noindent {\bf{Acknowledgments}}: We would like to
express our heartfelt thank and sincere gratitude to Professor M.V.
Takook for his helpful discussions. We also thank the referees for
their useful comments and suggestions.

\begin{appendix}
\section{Appendix}
\catcode `\@=11 \@addtoreset{equation}{section}
\def\theequation{\Alph{section}.\arabic{equation}}
\catcode `\@=12 \catcode `\@=11 \@addtoreset{proposition}{section}
\def\theproposition{\arabic{section}.\arabic{proposition}}
\catcode `\@=12 \catcode `\@=11 \@addtoreset{theorem}{section}
\def\thetheorem{\arabic{section}.\arabic{theorem}}
\catcode `\@=12 \catcode `\@=11 \label{A}

\subsection{The field equation (\ref{3.13})}

In this Appendix the field equation (\ref{3.13}) is considered:\b
\tilde{D}^\delta_\alpha\Psi_\delta=0. \label{A.89}\e The solutions
of equation (\ref{A.89}) obviously satisfied the field equation
(\ref{dirac1}). This equation can be written as a simple form \b
\not x\not\partial^\top\Psi_\alpha(x)-x_\alpha\not
x\not\Psi-3\Psi_\alpha(x)=0 .\e  For prettiness it can be written as
\b \left(Trpr\not x\not\partial^\top-3\right)\Psi_\alpha(x)=0 ,
\label{A.3} \e Now we rewrite the equation (\ref{A.3}) as\b
\left(Trpr\not
x\not\partial^\top-3\right)\Psi_\alpha(x)-\partial^\top_\alpha\not
x\not\Psi(x)=0 ,\e which is invariant under the gauge
transformation: \b \Psi_\alpha (x) \longrightarrow \Psi_\alpha'
(x)=\Psi_\alpha (x)+ D_{\frac{3}{2}\alpha}\zeta ,\label{A.5}\e where
$\zeta$ is an arbitrary spinor field. Let us introduce a gauge
fixing parameter $\lambda$ for this case. Now the wave equation
reads as \b \left(Trpr\not
x\not\partial^\top-3\right)\Psi_\alpha(x)-\lambda\partial^\top_\alpha\not
x\not\Psi(x)=0 .\label{A.6}\e The above equation under the gauge
transformation (\ref{A.5}) becomes \b \left(Trpr\not
x\not\partial^\top-3\right)\Psi_\alpha(x)-\lambda\partial^\top_\alpha\not
x\not\Psi(x)=(\lambda-1)\partial^\top_\alpha\not x\gamma^\beta\
D_{\frac{3}{2}\beta}\zeta. \e With the choice of value $\lambda\neq
1$, the gauge field $D_{\frac{3}{2}\alpha}\zeta$ is solved to the
field equation only if $\zeta$ obeys \b \partial^\top_\alpha\not
x\gamma^\beta D_{\frac{3}{2}\beta}\zeta=(Trpr\not
x\not\partial^\top-3)D_{\frac{3}{2}\alpha}\zeta=0,\e that means the
gauge field $\zeta$ is fixed.
 The adjoint field $\bar\Psi_\alpha$
satisfied the equation:
\b\bar\Psi_\alpha\gamma^4\overleftarrow{\not\partial}^\top\not
x\gamma^4-\bar\Psi_\beta\gamma^4(\gamma^\beta x_\alpha\not
x)\gamma^4+3\bar\Psi_\alpha-\lambda\bar\Psi_\beta\gamma^4\gamma^\beta\not
x\overleftarrow{\partial}_\alpha^\top\gamma^4=0. \label{3.37}\e By
putting $\Psi_\alpha$ in equation (\ref{A.6}), we find that the
spinor fields $\psi_1,\psi_2$, and $\psi_3$ must obey the following
equations: \b (\not x\not\partial^\top-3)\psi_1=0,\label{A.10}\e
\b(\lambda-1)[4\not
x+\not\partial^\top]\psi_2+2(1-2\lambda)\psi_3=\lambda\not
Z^\top\psi_1,\label{A.11}\e \b(\lambda-1)[4\not
x+\not\partial^\top]\psi_2+[\not
x\not\partial^\top-2(2\lambda+1)]\psi_3=(\not xx.Z+\lambda\not
Z^\top)\psi_1.\label{A.12}\e By multiplying equation (\ref{A.10}) in
$\not
\partial^\top$, we have
$$Q_0\psi_1=0.$$
For $\lambda=\frac{1}{2}$ equation (\ref{A.11}) becomes\b\not
\partial^\top\psi_2+4\not x\psi_2=-\not
Z^\top\psi_1,\label{A.13}\e   and equation (\ref{A.12}) by using the equation (\ref{A.13})
becomes \b\not x\not
\partial^\top\psi_3-4\psi_3=\not xx\cdot Z\psi_1.\label{A.14}\e  Using equation (\ref{A.13})
and multiplying it in $\not
\partial^\top$ , we
obtain\b Q_0\psi_2=-4\psi_2-\not Z^\top\not
x\psi_1+2Z\cdot\partial^\top\psi_1+4Z\cdot x\psi_1 ,\e
 and by using equation (\ref{A.14}), we
obtain\b \not xQ_0\psi_3=-4\not x\psi_3+2x\cdot Z\psi_1+\not
Z^\top\not x\psi_1 .\e Since the homogenized coefficient is the same
for the functions $\psi_1$,$\psi_2$, and $\psi_3$ therefor
$$Q_0\psi_2=0,$$ $$Q_0\psi_3=0.$$ Then we
have \b\psi_2=\frac{1}{4}(-\not Z^\top\not
x\psi_1+2Z\cdot\partial^\top\psi_1+4x\cdot Z\psi_1),\e \b\not
x\psi_3=\frac{1}{4}(2x\cdot Z\psi_1+\not Z^\top\not x\psi_1).\e So
the solution be written as follow:
$$\Psi_\alpha=(-\frac{1}{4}Z^\top_\alpha-\frac{5}{4}\gamma^\top_\alpha\not
xx\cdot Z+\frac{1}{4}\not Z^\top\not
x\partial^\top_\alpha-\frac{1}{2}\partial^\top_\alpha
Z\cdot\partial^\top$$ \b- x\cdot
Z\partial^\top_\alpha-\frac{1}{2}\gamma^\top_\alpha\not x
Z\cdot\partial^\top+\frac{1}{4}\not
Z^\top\gamma^\top_\alpha)\psi_1.\e

\subsection{The proof of equation (\ref{4.13})}

By replacing equation (\ref{4.12}) in equation (\ref{4.3}) we  have\b \not
x\not\partial^\top\psi_3+\not x(\not
x\not\partial^\top-4)\left(\frac{1}{2}\not Z^\top\not
x\psi_1+\psi_g\right) =\not x x.Z\psi_1.\label{A.108}\e Multiplying
equation (\ref{A.108}) in $\not x$, we can get:\b
\not\partial^\top\psi_3+\frac{1}{2}\not x\not\partial^\top\not
Z^\top\not x\psi_1-2\not Z^\top\not x\psi_1-4\psi_g= x.Z\psi_1 .\e
By using the equation:\b \not\partial^\top\not Z^\top\not
x\psi_1=4\not x x.Z\psi_1-2\not Z^\top\psi_1+2\not x
Z.\partial^\top\psi_1,\e and the following conditions
$$ x.Z\psi_1=-\not Z^\top\not x\psi_1 \:\:\:\: , \:\:\:\:
\not\partial^\top\psi_g=0,$$ we
have\b\not\partial^\top\psi_3=4\psi_g.\e In finally $\psi_3$
obtained as follows\b \psi_3=\not x \psi_g.\e

\subsection{Two-point function}

Here, the two-point function is calculated with respect to $x'$,
which satisfies equation (\ref{3.24}). In this case, we obtain
 \b S_1(x,x')\gamma^4(\overleftarrow{\not\partial'^\top}\not
x-1)\gamma^4=0,\e \b \gamma^\top_\alpha S_3(x,x')\gamma^4\overleftarrow{\partial'^\top}\not
x\gamma^4+D_{\frac{3}{2}\alpha}S_2(x,x')\gamma^4(\overleftarrow{\not\partial'^\top}\not x-4)\not
x\gamma^4=S_1(x,x')\gamma^4\not xx.\theta'\gamma^4 ,\e
$$ 2(1-2\lambda)\gamma^\top_\alpha S_3(x,x')+D_{\frac{3}{2}\alpha}S_2(x,x')\gamma^4\left[(1-\lambda)\overleftarrow{\not\partial'^\top}\not
x-2(1-2\lambda)\right]\not x\gamma^4$$\b=\lambda
S_1(x,x')\gamma^4\gamma^\top.\theta'\gamma^4.\e So, $S_2$, and $S_3$
are given by
$$ D_{\frac{3}{2}\alpha}S_2(x,x')=-S_1\frac{1}{2}\gamma^4x'.\theta\gamma^4-S_g ,
$$\b \gamma^\top_\alpha S_3(x,x')=-S_g{\gamma^4}\not x{\gamma^4} .\e Finally
the two-point function in this case (for $\lambda=\frac{1}{2}$) is
\begin{equation} S_{\alpha \alpha'}(x,x')=
S_1(x,x')\overleftarrow{D}_{\alpha \alpha'}(x,
\partial^\top ;x', {\partial'}^\top) ,\end{equation} where
$$
D_{\alpha \alpha'}(x,
\partial^\top ;x', {\partial'}^\top)= \frac{1}{2}\left[3\theta_\alpha \cdot
\theta'_{\alpha'}+\overleftarrow{\not\partial'^\top}x\cdot\theta'_{\alpha}
+\gamma^4\gamma^\top_{\alpha'}\gamma\cdot\theta\gamma^4\right].$$

 \end{appendix}

\end{document}